# Circuit-Level Modeling for Concurrent Testing of Operational Defects due to Gate Oxide Breakdown

Jonathan R. Carter, Sule Ozev, and Daniel J. Sorin

Department of Electrical and Computer Engineering, Duke University

## Abstract

*As device sizes shrink and current densities increase, the probability of device failures due to gate oxide breakdown (OBD) also increases. To provide designs that are tolerant to such failures, we must investigate and understand the manifestations of this physical phenomenon at the circuit and system level. In this paper, we develop a model for operational OBD defects, and we explore how to test for faults due to OBD. For a NAND gate, we derive the necessary input conditions that excite and detect errors due to OBD defects at the gate level. We show that traditional pattern generators fail to exercise all of these defects. Finally, we show that these test patterns can be propagated and justified for a combinational circuit in a manner similar to traditional ATPG.*

## 1 Introduction

With decreasing feature dimensions and increasing transistor counts in typical designs, the reliability of semiconductor devices is an increasing concern. One emerging reliability problem is gate oxide breakdown (OBD). The technology trend towards decreasing gate oxide thicknesses results in a greater probability of oxide degradation and breakdown during the lifetime of the devices.due.

Most fault tolerance techniques, such as error correcting codes or high level redundancy, work at the behavioral level without the assumption of any fault model. However, this generality typically imposes high overheads in terms of hardware, performance, and power consumption. Circuit level fault tolerance methods, such as dynamic memory repair, on the other hand, rely on traditional fault models, such as the stuck-at fault model, or coupling fault model. Thus, they may be incapable of detecting errors due to operational defects that do not manifest themselves in traditional ways. To provide fault tolerance at the circuit level, we must investigate the underlying physical defects and understand their manifestations at the circuit level.

The OBD phenomenon has been studied extensively at the device level, yet little research has explored OBD's circuit level implications. It may appear that new fault models are not needed, since oxide breakdown does not result in logic errors at its early stages of soft breakdown, and it results in a simple stuck-at logic error when the final hard breakdown happens. However, OBD affects the dynamic behavior of the circuit long before the final hard breakdown, starting from the early stages of the process. Soft OBD causes much longer delays than expected for the circuit and can thus introduce logic errors. Despite the manifestation of logic errors, traditional fault models may not predict the behavior of such defects accurately. Section 2 discusses related previous work in fault modeling, and it focuses on existing circuit-level fault modeling and the recent device-level modeling of OBD.

In the context of operational defects due to OBD, we address the issues of fault modeling and testing. We show that, for both issues, OBD operational defects demand careful circuit-level consideration. Our contributions in these two areas are the following:

- Fault modeling (Section 3): We show that circuit behavior in the presence of OBD defects exhibits similar aspects to general transition or delay faults, but differs in important ways. Based on published data at the device level, we develop a circuit-level model that closely matches the underlying physical phenomenon.

- Concurrent testing (Section 4): We show that OBD defects require more sophisticated approaches for fault excitation and detection. Particularly, the window of opportunity for detecting OBD defects before they reach their final hard breakdown status depends on the latency of the detection mechanism and the delay caused by these defects. We develop the necessary input conditions for a NAND gate and infer conclusions for other gates. We show that the derived input conditions and output behavior can be propagated to the primary inputs and outputs of a combinational circuit in a manner similar to traditional ATPG.

## 2 Related Work

**Physical Causes of Faults.** The reliability of electronic devices under discrete environmental stress, such as radiation [20], and continuous functional stress due to applied electric fields [14, 22, 2] has been studied since the early days of semiconductor manufacturing. Extensive research has been conducted on the failure-causing physical phe-nomena, such as electromigration (EM) [22, 2] and OBD [6]. The effect of OBD at the circuit level is a shift in the LOW and HIGH voltage values and eventually bit errors for CMOS devices [21, 15, 16].

OBD defects have been studied in the past decade, although this research has focused on hard breakdowns (also known as gate oxide shorts or GOS). Segura et al. manufacture circuits with intentional hard OBD defects and measure their characteristics [19]. They also study the distinct responses of NMOS and PMOS transistors to hard OBD defects [18], and they propose test patterns for $I_{DDQ}$ measurements to detect these defects. Renovell et al. [12] develop a transistor-level model for hard OBD defects that does not rely on splitting transistor dimensions. They also propose test patterns for detecting hard OBD defects through delay testing [13]. While hard OBD defects, which are typically tested after manufacturing, are identical to the final stage of the OBD defects that we analyze here, our goal is to create a model to help us develop concurrent test/ diagnose/repair mechanisms for OBD defects before final hard breakdown occurs. Since hard OBD can potentially harm the rest of the circuit, modeling OBD progression is essential to ensure fault-tolerant operation.

**Fault Modeling.** Several structural fault models have been developed for logic circuits and storage components over the past few decades [1]. The stuck-at fault model is the most commonly used model in VLSI testing and fault tolerant design. In this model, a defect manifests itself as a signal consistently having a certain value (zero or one) independent of the input. For example, an unintended short circuit between the two inputs of an XOR gate results in a stuck-at-zero fault at the output. The coupling fault model—in which a write to a certain memory location always prompts a write to a neighboring location or locations—has been defined for storage components [4].

The path delay fault model [10, 9] and the transition fault model [11, 5, 23, 17] have been used to test for manufacturing defects and process variations. In the path delay fault model, unexpectedly high process variations as well as point defects increase the overall delay of a path beyond the designated clock period. Paths with the highest logic depth are most likely to incur logic errors. Transition faults are based on point defects as opposed to the collective defects in the path delay fault model. In the transition fault model, resistive connections cause a circuit node to rise or fall too slowly. While both transition and path delay fault models are dynamic, they do not differentiate among the underlying physical defect locations and types. Thus, test generation methods targeting these faults are insensitive to which input combinations cause the desired transition at the fault location. For example, for a NAND gate, input transitions (11,00), (11,01), and (11,10) are identical in testing for a slow-to-rise transition fault at the output.

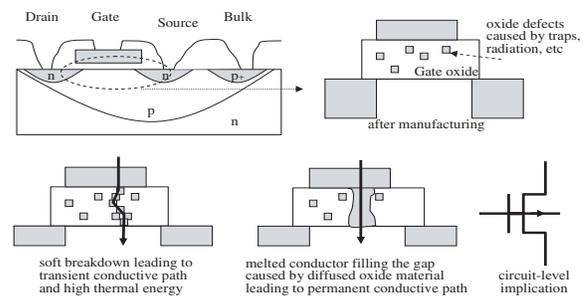

**Figure 1. Oxide breakdown process in transistor**

While the physical phenomena causing OBD have been studied in detail, limited work has been devoted to investigating the circuit level manifestations and modeling of these defects. Some recent work [16, 15] models the oxide breakdown phenomenon for an inverter using SPICE circuit elements. This work matches the model data to the measured data for the transfer characteristics of the inverter. Based on this circuit model, it is clear that oxide breakdown progressively impacts the circuit behavior and finally results in a logic error.

## 3  Gate Oxide Breakdown Defects

In this section, we describe the underlying physical phenomenon for OBD defects, our simulation model, and the dynamic behavior of circuits with OBD defects.

### 3.1  Underlying Physical Phenomenon

In a newly manufactured device, there is a random number of traps (imperfections) within the oxide. Over time, due to operational stress, more traps form, leading to small transient conductive paths within the oxide. During this process, current conduction is due to a combination of conductive paths formed by oxide traps and tunneling through the oxide. The formation of these conductive paths is referred to as soft breakdown (SBD). These paths may disappear after an initial injection of high current density that causes high temperature at the defect location. High temperature may relocate some of the oxide traps, breaking the conduction path. However, after a number of SBD incidents, many traps exist within the oxide, leading to a consistently high current density. Eventually, the resulting heat may generate holes in the oxide and melt the conductive material at the gate. Thus, a persistent low-resistance path is formed, leading to a hard breakdown (HBD). This progressive phenomenon is illustrated in Figure 1.

As discussed in Section 2, prior work [16] concluded that oxide breakdown has negligible effect on circuit timing and that the degraded logic value can be restored by the next logic stage. This argument is true if either the breakdown process is in its early stages or the source driving the breakdown gate has large current supply capability.

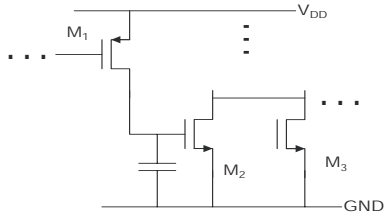

**Figure 2. Impact of OBD on upstream and downstream logic**

Since OBD is a progressive process, once soft breakdown starts, the leakage current starts continuously increasing, as experimentally shown in previous work [7, 8]. Allowing the defect to reach its final HBD stage, where it generates a stuck-at like fault, may potentially result in catastrophic impacts for upstream logic, as shown in Figure 2. Since a HBD for transistor M2 practically results in a short circuit between the gate and source/drain of the transistor, the drain of the driving transistor M1 will be effectively shorted to ground. This persistent connection to ground can potentially damage M1 and/or the supply. Thus, such defects need to be detected and repaired before HBD.

The progression of the breakdown may result in appreciable timing delays, creating logic errors if the defective transistor is driven by upstream logic. The timing delay is due to two factors, as illustrated in Figure 2. First, the leakage current through the breakdown path in M2 diverts a large portion of the current supplied by M1 to charge the gate capacitance of M2. Second, the degraded logic level at M2 results in a lower current for transistor M3, slowing down its operation as well. Thus, even if the HBD stage is not reached, the defect may cause logic errors.

### 3.2 Simulation Circuit Model

In order to understand the progression of OBD and how to detect OBD faults in embedded logic circuits during their operation, we developed a circuit model that can be used with HSPICE. As illustrated in Figure 3a, the oxide breakdown results in a resistive connection between the gate and the bulk of the device. This connection is followed by pn junctions to the source and the drain of the device. There is also a resistive connection between the breakdown location and the substrate. We assume that the substrate connection is farther away, resulting in a high resistance. Thus, the breakdown transistor can be modeled with two resistors and two diodes, as shown in Figure 3b.

With the diode-resistor based model illustrated in Figure 3b, the progression of OBD can be explained by an increase in the saturation currents of the diodes and a decrease in the resistance. This progression is mainly due to the high current density through the oxide causing high heat and generating fast deformations in the neighboring regions, resulting in an increase in the effective contact area between the gate and the bulk right underneath the

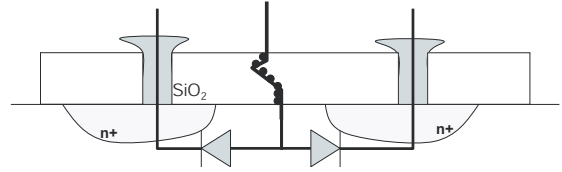

a. Device level model for OBD

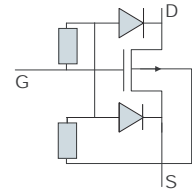

b. Simulation model for OBD

**Figure 3. Modeling the OBD Process**

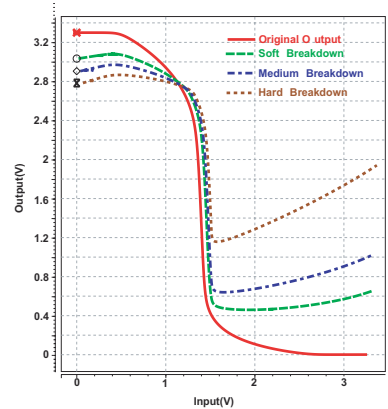

**Figure 4. Input/output characteristics for an inverter using the circuit model. Model explores OBD in the NMOS.**

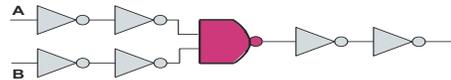

**Figure 5. Set-up used in HSPICE simulations**

gate. We have experimentally verified this model with respect to the measured data reported by Rodriguez et al. [16]. In terms of the voltage transfer characteristics of an inverter, the OBD in an NMOS transistor results in an upward shift in the VOL value, as illustrated in Figure 4. The OBD in a PMOS transistor results in a downward shift in the VOH value [16]. Considering only static voltage transfer characteristics is not sufficient. The timing effect of OBD results in logic errors long before the static characteristics start impacting the output.

### 3.3 Dynamic Behavior of Transistors with OBD

OBD in an internal transistor of a gate results in a transition delay at its output. However, the manifestation of this defect is different from the traditional transition fault model since the input conditions play an important role in its excitation. In HSPICE, to simulate the OBD effect for a

Proceedings of the Design, Automation and Test in Europe Conference and Exhibition (DATE'05)
1530-1591/05 $ 20.00 IEEE

**TABLE 1: NMOS and PMOS OBD progression in terms of transition delays for the set-up in Figure 5**

|  | | (01,11) | | (10,11) | | | (11,10) | | (11,01) | |
|---|---|---|---|---|---|---|---|---|---|---|
|  | $I_{sat}$, R | $N_A$ | $N_B$ | $N_A$ | $N_B$ | $I_{sat}$, R | $P_A$ | $P_B$ | $P_A$ | $P_B$ |
| Fault Free | 1e-30,10k | 96ps | 96ps | 96ps | 96ps | 1e-30,10k | 110ps | 110ps | 110ps | 110ps |
| MBD1 | 2e-28,500 | 118ps | 118ps | 118ps | 118ps | 1e-29,1k | 110ps | 360ps | 360ps | 110ps |
| MBD2 | 1e-27,100 | 156ps | 143ps | 144ps | 156ps | 1.1e-29,900 | 110ps | 736ps | 740ps | 110ps |
| MBD3 | 5e-27,20 | 190ps | 228ps | 230ps | 190ps | 1.2e-29,830 | 110ps | sa-0 | sa-0 | 110ps |
| HBD | 2e-24,0.05 | sa-1 | sa-1 | sa-1 | sa-1 | N/A | N/A | N/A | N/A | N/A |

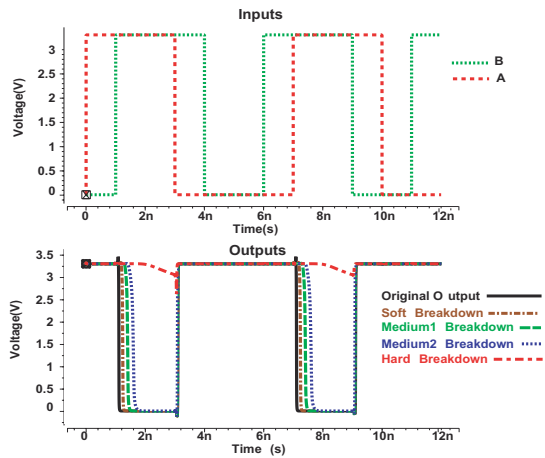

**Figure 6.** Progression of NMOS OBD for NAND

gate that is embedded in a logic path, we need to ensure that the defective gate is driven by other gates rather than by a voltage source (as in previous work). Thus, we employ the simulation set-up shown in Figure 5 for a NAND gate.

As explained in the previous sections, two events play a role in the transition delay of a gate under OBD breakdown. First, the leakage current through the oxide impedes the charging or discharging of the output node. Second, the lower swing of the output node due to the breakdown slows down the downstream logic driven by the defective gate.

Since OBD is a progressive defect, there is a window of opportunity in which to detect the errors caused by it. We would like to detect these defects long before the final HBD status is reached, since there is a potential danger of harming other components at this stage. The window of opportunity to detect the OBD defects is between the SBD stage and HBD stage as explained in Section 3.1. We refer to the state of the breakdown process within this time as medium breakdown (MBD).

Based on data from Linder et al. [7], the time between the first SBD incident and the final HBD is roughly 27 hours for a PFET with a 15Å oxide thickness. However, the detectability of an initial SBD defect is quite low since the delay caused by it can be transient and/or small. Thus, the window for detection/repair is really much shorter than 27 hours. The increase in the leakage current between SBD and HBD is exponential. Table 1 shows the saturation currents ($I_{sat}$) and resistance (R) values used in our model and the resulting transition delays for the NAND gate. The "Fault Free" row refers to the case with no OBD. Input sequences are shown in the first row of the table. The columns corresponding to entries $N_y$ ($P_y$) indicate an NMOS (PMOS) OBD in the transistor connected to input y.

Figure 6 shows the impact of an NMOS OBD on the timing of the NAND gate in Figure 5, assuming the defective NMOS transistor is connected to input A (i.e., $N_A$). We observe from Figure 6 that breakdown in the NMOS transistor causes a transition fault at the output of the gate that is independent of which input switches. The reason for this behavior is that the OBD defect mainly injects current into the drain of the transistor and limits the overall discharge current through the series connection regardless of which input switches. The same effect can be observed when the defective transistor is connected to input B.

## 4 Concurrent Testing for OBD Defects

The experiments in Section 3 show that transition faults caused by OBD can be detected only through a timing sensitive testing method. However, to excite the faults caused by OBD defects, specific input transitions are necessary.

### 4.1 Input Conditions for Detecting OBD Defects

While the NMOS OBD defects can be detected through any combination of input switching that results in a high-to-low switching at the output, this is not the case for the PMOS defects. Since the PMOS transistors are in parallel, the defective transistor can only cause a timing delay at the output if it is the only transistor charging the output node. Thus, PMOS defects in a NAND gate can be detected only through specific input combinations that result in a low-to-high transition at the output node.

The input-specific nature of PMOS OBD defect manifestation can be observed from Figure 7. The defective PMOS transistor in this case is connected to the input A. While seemingly all PMOS OBD defects result in a slow-to-rise transition fault, an OBD defect in the PMOS transistor that is connected to input A can only be detected if input A is switched from 1 to 0 while input B is kept at 1. This input-specific excitation of the OBD defect distinguishes it from the general transition faults that consider the rise and fall of intermediate circuit nodes independent of the transistor behavior. Similarly, an OBD defect in the PMOS transistor that is connected to input B can only be detected if input A is held at 1 and B is switched from 1 to 0.

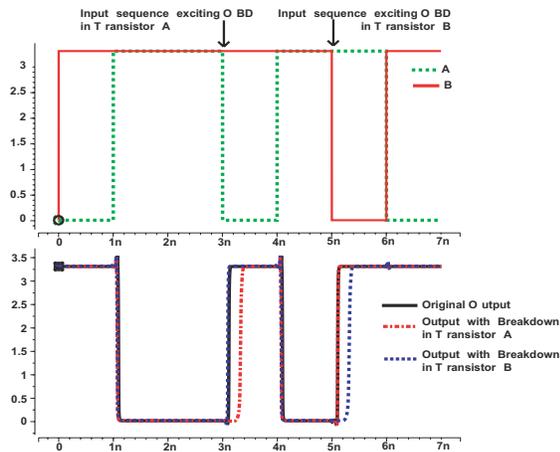

**Figure 7. Input-specific detection of transition faults caused by PMOS OBD defects**

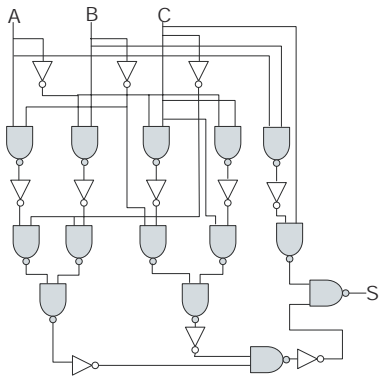

**Figure 8. Full adder circuit used as example**

To detect all OBD defects in a NAND, one of the input sequences {(10,11), (00,11), (01,11)} and the sequences {(11,10)} and {(11,01)} are necessary and sufficient.

### 4.2 Propagation of fault effects

Once the excitation conditions are met, an OBD defect results in a transition delay at the output of the defective gate. This transition delay can be propagated to the primary output through other gates in a similar fashion to traditional fault models, such as the stuck-at fault model. However, the detection of this fault may necessitate output capture earlier than the designated clock frequency of the digital circuit. This problem is similar to transition fault testing and several early capture methods have been developed [17].

It can be argued that the timing delay due to an OBD defect will eventually generate detectable logic errors at the output independent of the slack in the clock period due to the progressive nature of these defects. However, as experimentally shown [7, 3], the progression of OBD defects from the onset of appreciable leakage current is exponential. Thus, defects must be identified as soon as appreciable leakage current starts flowing. This window of opportunity depends on the timing slack in the detection mechanism. The diode-resistor based model can be used to determine the delay caused by the OBD defects at various stages of the progression. This information helps the scheduling of test/diagnosis/repair intervals of fault-tolerance schemes.

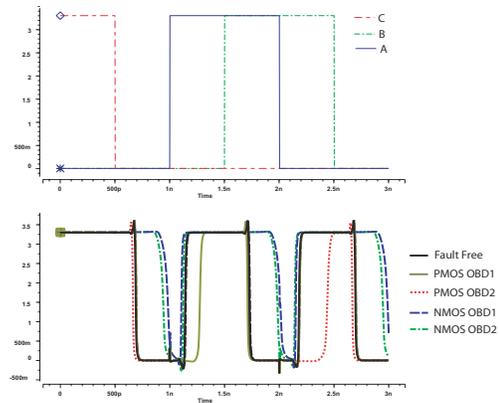

**Figure 9. Propagation of transition fault effects for PMOS and NMOS OBD defects**

### 4.3 Illustrative example: Sum output of full adder

In this section, we show that the manifestations of OBD defects can be propagated through a logic path and can be observed at the output through a timing delay, even though the degraded logic value is restored in the end. As an experimental circuit, we have used a combinational circuit: the sum bit of a full adder, implemented using 14 NAND gates and 11 inverters, as shown in Figure 8. For illustrative purposes, we implemented this logic function without any optimizations, resulting in a logic depth of 9.

We injected single OBD errors into the NAND gate that has four stages in both the upstream and downstream logic. Figure 9 shows the inputs and outputs of this circuit for the injected faults. The necessary input sequences for the NAND gate have been propagated to the primary inputs, A, B, and C. The delays due to the OBD defects in the four transistors inside the NAND gate (injected one at a time) can be observed at the primary output, the sum bit.

In the example circuit, there are 56 distinct locations for OBD defects in the 14 NAND gates. Some of these faults are undetectable due to the intentional redundancy in the circuit. 18 out of 72 input transitions are necessary and sufficient to detect the 32 testable OBD faults in the circuit.

## 5 Applicability to General Circuits

Our analysis for the NAND gate can be generalized as follows. The OBD breakdown of a transistor can be detected at an output node only if that transistor is excited at the switching of the output node and if no other transistor that is connected to the defective transistor in parallel is excited. As an example, for a traditional NOR gate, one of the input combinations {(10,00), (01,00), (11,00)}, and the input sequences {(00,01)}, and {(00,10)} are necessary and sufficient to detect all four OBD defects.

Testing for OBD defects requires transitional inputs. For combinational circuits, test pattern generation (TPG) for OBD defects is of the same computational complexity as for stuck-at faults. Similar to transition faults, sequential TPG for OBD defects is more complicated than sequential TPG for stuck-at faults due to the need to generate two distinct input combinations at consecutive clock cycles. Thus, we need design-for-testability methods to enhance controllability and/or observability of circuit nodes.

In the full adder example, the small set of input transitions (18 out of the 72 possible) that suffices to detect all OBD faults makes built-in-testing for such defects promising, particularly for safety-critical applications.

For a traditional NAND gate, the input sequences that detect all the internal intra-gate EM faults are {(11,01)}, {(11,10)}, and {(01,11), (10,11), (00,11)}. At a first glance it may seem that the test inputs that target internal EM defects also detect all OBD defects in a logic gate. However, due to the current injecting nature of OBD defects, this may not always be true, especially for complex gates and high performance logic. Therefore, there is a need to use the circuit models for OBD defects in order to generate test input conditions to detect errors caused by them.

## 6 Conclusions

Traditional fault modeling techniques that focus only on static behavior or only on the transition of inter-gate nodes may fail to target emerging defects due to gate oxide breakdown. Detecting the faults caused by such defects is particularly important for safety critical applications and applications that need to run autonomically for long periods of time. In this study, we have used a NAND gate to show that to target such defects, the circuit-level implications need to be considered for test generation.

**Acknowledgments**

This material is based upon work supported by the National Science Foundation under Grant CCR-0309164, the National Aeronautics and Space Administration under Grant NNG04GQ06G, and a Duke Warren Faculty Scholarship (Sorin).